\documentclass[cits]{PoS}

\title{Adiabatic expansion and magnetic fields in AGN jets}

\ShortTitle{Adiabatic expansion and magnetic fields in AGN jets}

\author{\speaker{Alexander~B.~Pushkarev}%
         \thanks{{\bf Acknowledgements.} 
         Y.Y.~Kovalev is a Research Fellow of the Alexander von Humboldt Foundation.
         The National Radio Astronomy Observatory is a facility of the National Science Foundation operated
         under cooperative agreement by Associated Universities, Inc. This research has made use of the
         NASA/IPAC Extragalactic Database (NED) which is operated by the Jet Propulsion Laboratory,
         California Institute of Technology, under contract with National Aeronautics and Space
         Administration. We would like to thank Matt Lister and Tuomas Savolainen for useful comments and discussion.
}\\
        Max-Planck-Institut f\"ur Radioastronomie, Auf dem H\"ugel 69, 53123 Bonn, Germany \\
        Pulkovo Observatory, Pulkovskoe Chaussee 65/1, 196140 St. Petersburg, Russia\\
        Crimean Astrophysical Observatory, 98409 Nauchny, Crimea, Ukraine\\
        E-mail: \email{apushkar@mpifr-bonn.mpg.de}}

\author{Yuri~Y.~Kovalev\\
        Max-Planck-Institut f\"ur Radioastronomie, Auf dem H\"ugel 69, 53123 Bonn, Germany\\
        Astro Space Center of Lebedev Physical Institute, Profsoyuznaya 84/32, 117997 Moscow, Russia\\
        E-mail: \email{ykovalev@mpifr-bonn.mpg.de}}

\author{Andrei~P.~Lobanov\\
        Max-Planck-Institut f\"ur Radioastronomie, Auf dem H\"ugel 69, 53123 Bonn, Germany\\
        E-mail: \email{alobanov@mpifr-bonn.mpg.de}}

\abstract{
Results of high-resolution simultaneous multi-frequency 8.1-15.4~GHz
VLBA polarimetric observations of relativistic jets in active galactic
nuclei (the MOJAVE-2 project) are analyzed. We compare characteristics
of VLBI features with jet model predictions and test if adiabatic
expansion is a dominating mechanism for the evolution of relativistic
shocks in parsec-scale AGN jets. We also discuss magnetic field
configuration, both predicted by the model and deduced from electric
vector position angle measurements.
}

\FullConference{The 9th European VLBI Network Symposium on The role of VLBI in the
Golden Age for Radio Astronomy and EVN Users Meeting\\
                 September 23-26, 2008\\
                 Bologna, Italy}

\begin{document}

\section{Introduction}

In spite of the fact that our understanding of active galactic nuclei phenomenon has been significantly 
improved over the past decade \cite{Lobanov06}, there are many aspects of nature of the relativistic
outflows (jets), like their composition, formation, acceleration, collimation, etc, which remain to be
strongly debated. VLBI observations provide us with the unique tool to explore the jets in AGN
with milliarcsecond angular resolution corresponding to parsec-scale resolution in linear size. 
Typically, parsec-scale flows are characterized by one-sided knotty structure with sometimes pronounced 
curvature, rapid variations of flux density (e.g. \cite{Wagner95,Savolainen08}), and detection of 
superluminal motions \cite{Kellermann04}. 

Relativistic shocks propagating down the jets are expected to be prominent on these scales,
which is confirmed by detection of strong polarization \cite{Pushkarev05,Ros00} and 
enhancement of the magnetic field in the brightest jet features (from the turnover frequency images)
\cite{Lobanov97}. The general evolution of a shock-induced flare in a jet is considered by \cite{Marscher96}
to pass through the Compton, synchrotron, and adiabatic stages, referring to the dominant energy loss 
mechanism. In the current paper we present recent results of testing adiabatic expansion of relativistic 
shocks in several active galactic nuclei.

Throughout the paper the $\Lambda$CDM cosmological model with $H_0=70$~km~s$^{-1}$~Mpc$^{-1}$, $\Omega_m=0.3$,
and $\Omega_\Lambda=0.7$ is adopted. Spectral index $\alpha$ is defined as $S\propto\nu^\alpha$.

\section{Observational data in use}

We use experimental data collected and processed in the framework of the 
MOJAVE-2\footnote{{\bf M}onitoring {\bf O}f {\bf J}ets in {\bf A}ctive galactic nuclei with 
{\bf V}LBA {\bf E}xperiments, see {\tt http://www.physics.purdue.edu/MOJAVE/}} project. 
The observations were performed by the MOJAVE team in 2006 at VLBA, a VLBI network of ten 25-m
telescopes, in a dual-polarization mode, quasi-simultaneously at four frequencies of 8.1, 8.4, 
12.1, \& 15.4~GHz. The scan lengths were chosen to achieve roughly the same image rms at each 
observing frequency. The observed sample consists of 192 sources, see for details \cite{Lister08}.

\section{Results and Discussion}

Using the shock-in-jet-model \cite{Marscher90} we have investigated an evolutionary scenario in which 
a bright jet feature is treated as an independent, geometrically thin relativistic shock developing in a flow
with the emission dominated by adiabatic energy losses. It is also assumed that the jet plasma has
a power-law particle energy distribution $N(E)\,d\,E\propto E^{-s}\,d\,E$.
Following \cite{Baum97} and applying Gauss's flux theorem and Lorentz transformations to the cross section
of the jet, the magnetic 
field can be approximated as $B\propto d^{-n}$, where $d$ is the transverse jet size and $n$ describes the 
orientation of the magnetic field ($n=1$ for the transverse field and $n=2$ for the longitudinal field). 
Assuming non-changing Doppler factor along the jet, the model brightness temperature of each 
jet component, $T_{\rm b,\,jet}$, can be related, as it was shown in \cite{Lobanov00}, to the measured 
brightness temperature of the core, $T_{\rm b,\,core}$, as 
$T_{\rm b,\,jet}=T_{\rm b,\,core}(d_{\rm jet}/d_{\rm core})^{-\xi}$, where $d$ represents the measured 
sizes of the core and jet components, and $\xi=n+1-\alpha(n+4/3)$.

We compare the measured brightness temperatures with those predicted by the model,
determining the spectral index of each jet component as well as the B-field orientation 
(if polarization emission is detected, preferably at 15.4~GHz, by rotating electric vectors by $90^\circ$
due to the optically thin synchrotron radiation detected in the jets).

In this paper we discuss the results for two sources, 1128$-$047 and 2155$-$152, having prominent outflows
with sufficient number of jet components suitable for the analysis.

\subsection{1128$-$047}

\begin{figure}[b]
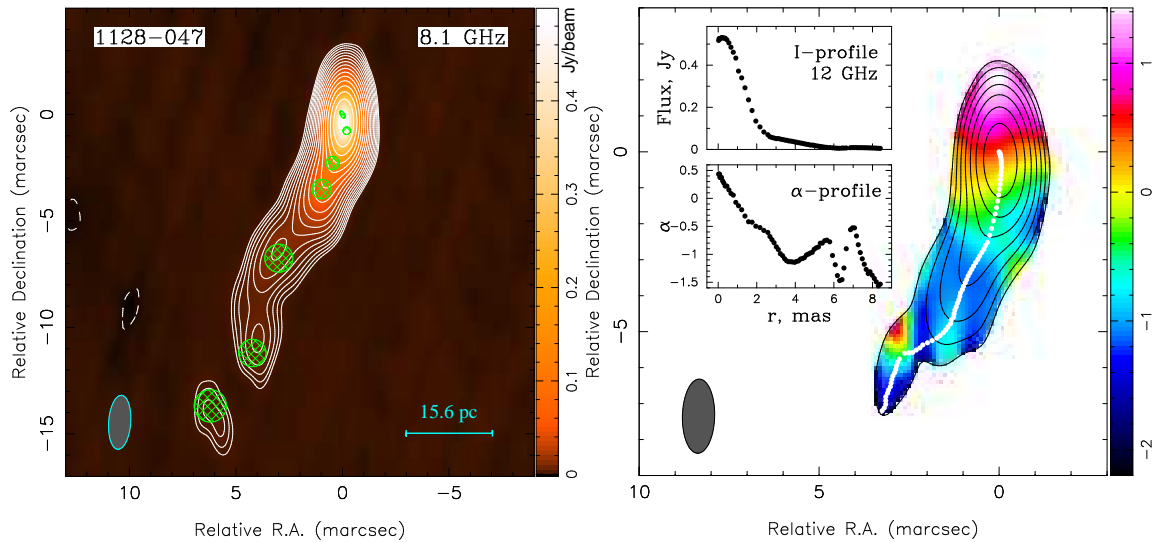

\includegraphics[height=.5\textwidth,angle=-90,clip=true]{figure1a.eps}
\includegraphics[height=.5\textwidth,angle=-90,clip=true]{figure1b.eps}
\caption{
{\it Left panel:} Total intensity image of radio galaxy 1128$-$047 at 8.1~GHz at epoch of 2006.92 
with the model of 7 Gaussian components superimposed. The lowest contour is plotted at 0.35\% of the peak 
brightness of 501~mJy$/$beam. Shaded ellipse represents the FWHM of the restoring beams of 
$2.54\times1.04$~mas at ${\rm PA}= -4.0^\circ$. 
{\it Right panel:} Spectral index distribution in 1128$-$047 calculated between 8.1 and 12.1~GHz 
with the 12.1~GHz total intensity contours overlayed. White dots represent the total intensity ridge 
line along which we plot the profiles of total intensity and spectral index as an inset image. 
The lowest contour is plotted at 0.84\% of the peak brightness of 531~mJy$/$beam. Shaded ellipse 
represents the FWHM of the restoring beam of $2.06\times0.89$~mas at ${\rm PA}= -1.4^\circ$.
}
\label{fig1}
\end{figure}

This flat-spectrum radio galaxy with redshift $z=0.266$ \,(3.91~pc$/$mas) and mass of the central object
of $5.2\times10^6\,{\rm M}_\odot$ \cite{Woo02} shows one-sided jet structure (Fig.~\ref{fig1}, left) 
which was fitted by seven distinct Gaussian components at 8.1~GHz. The model fitting was performed 
in DIFMAP \cite{Shepherd97}. 
In case of applying an elliptical Gaussian component for the VLBI core the minor axis of the ellipse
has been used to estimate $d$, since the VLBI core component has a strong tendency to be elongated in 
the innermost jet direction \cite{Kovalev05}.

\begin{figure}[t]
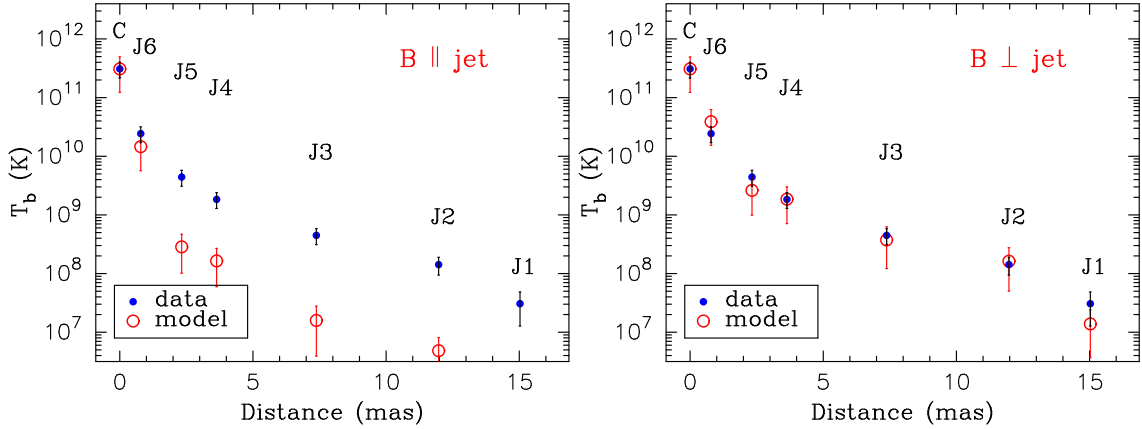

\includegraphics[height=.495\textwidth,angle=-90,clip=true]{figure2a.eps}
\includegraphics[height=.495\textwidth,angle=-90,clip=true]{figure2b.eps}
\caption{
{\it Left panel:}
Brightness temperature changes along the jet in 1128$-$047 at 8.1~GHz. Filled blue circles
are the measured values. Open red circles are the predicted brightness temperatures from
the shock-in-jet model with adiabatic losses dominating the radio emission assuming predominantly
longitudinal magnetic field orientation to the local jet direction. The errors are given at $3\sigma$ level.
{\it Right panel:} The same data as on the left but assuming transverse magnetic field orientation.
Big discrepancies (left) and remarkable agreement (right) between measured and model brightness
temperature values lead to the conclusion of perpendicular orientation of the magnetic field in the jet.
}
\label{fig2}
\end{figure}

Using the VLBI images at 8.1 and 12.1~GHz, we constructed spectral index distribution and its slice 
along the jet ridge line (Fig.~\ref{fig1}, right). Since the linear polarization in this source has 
not been detected down to about 1~mJy level, we calculated the model brightness 
temperatures for all of the jet components assuming (i) parallel (Fig.~\ref{fig2}, left) and 
(ii) perpendicular  magnetic filed configuration (Fig.~\ref{fig2}, right). In the latter case, the values 
of $T^{\rm obs}_{\rm\,\,\,b}$ and $T^{\rm mod}_{\rm\,\,\,b}$ agree within the errors that allows us to 
suggest predominantly transverse orientation of magnetic field in the jet.

\subsection{2155$-$152}

This flat-spectrum source with redshift $z=0.672$ \,(7.03~pc$/$mas) and mass of the central object of 
$3.9\times10^7\,{\rm M}_\odot$ \cite{Woo02} was classified as a BL Lac object by Craine et al. 
\cite{Craine76} and then as a quasar by V\'eron-Cetty \& V\'eron \cite{Veron06} on the basis of its optical 
spectrum. 

\begin{figure}[b]
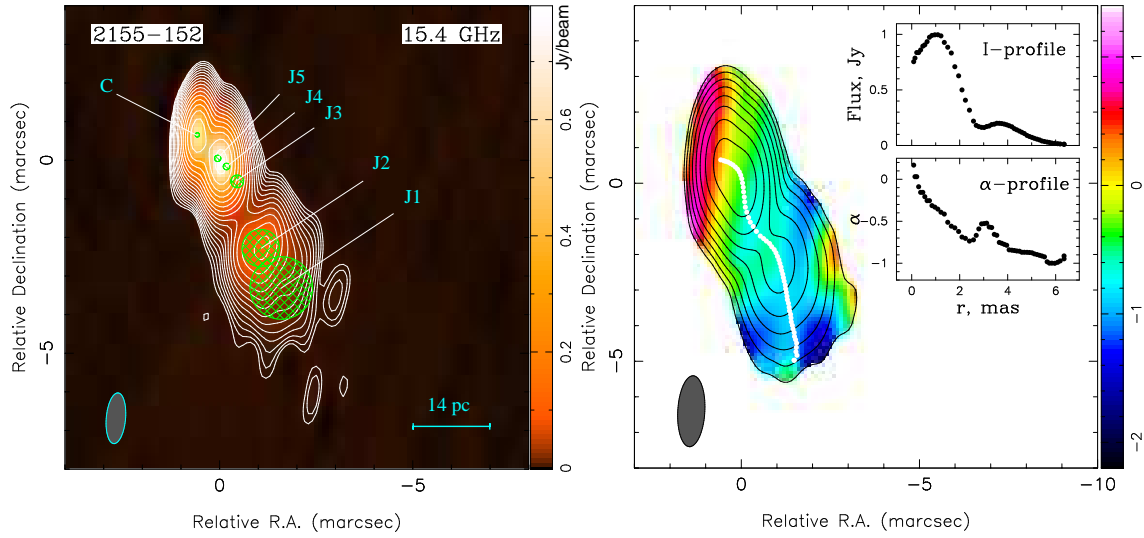

\includegraphics[height=0.495\textwidth,angle=-90,clip=true]{figure3a.eps}
\includegraphics[height=.495\textwidth,angle=-90,clip=true]{figure3b.eps}
\caption{
{\it Left panel:} Total intensity image of quasar 2155$-$152 at 15.4~GHz at epoch 2006.92 with 
model fitted Gaussian components superimposed. The lowest contour is plotted at 0.24\% of 
the peak brightness of 796~mJy$/$beam. Shaded ellipse represents the FWHM of the restoring beam of 
$1.32\times0.50$~mas at ${\rm PA}= -5.0^\circ$.
{\it Right panel:} Spectral index distribution in 2155$-$152 calculated between 8.1 and 15.4~GHz
with the 15.4~GHz total intensity contours overlayed. White dots represent the total intensity ridge 
line along which we plot the profiles of total intensity and spectral index as an inset image.
The lowest contour is plotted at 0.45\% of the peak brightness of 1000~mJy$/$beam. 
Shaded ellipse represents the FWHM of the restoring beam of $1.99\times0.76$~mas at ${\rm PA}= -3.2^\circ$. 
}
\label{fig3}
\end{figure}

\begin{figure}[t]
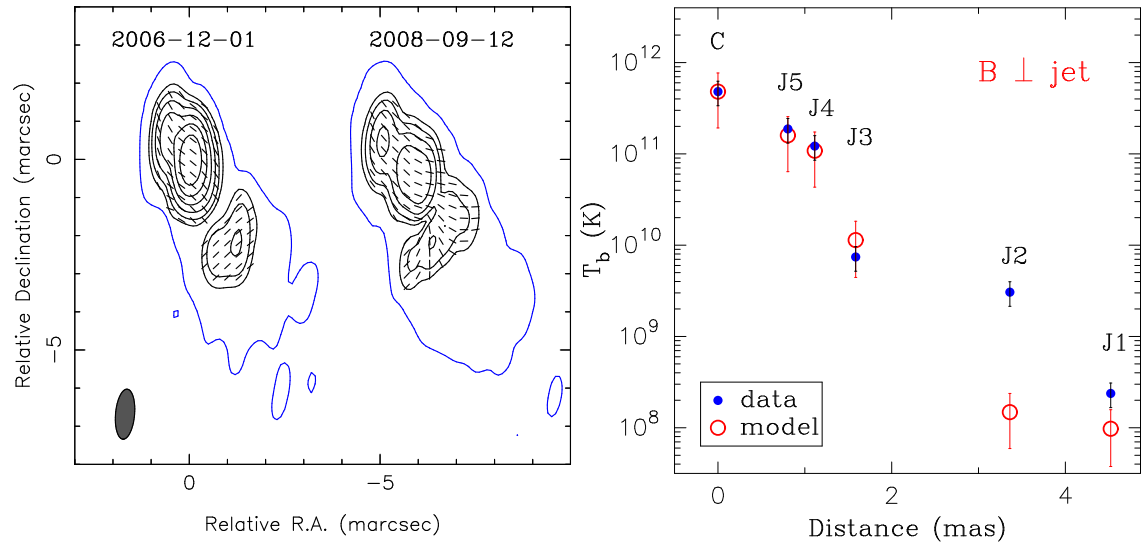

\includegraphics[height=.495\textwidth,angle=-90,clip=true]{figure4a.eps}
\includegraphics[height=.495\textwidth,angle=-90,clip=true]{figure4b.eps}
\caption{
{\it Left panel:} Linear polarization images of 2155$-$152 at 15.4 GHz with electric vectors superimposed on 
epochs 2006.92 and 2008.70. The bottom P-contour is plotted at 1.8~mJy for both epochs. The blue contour represents 
the total intensity bottom contour at a level of 1.9~mJy for both epochs. The sheath-spine structure with transverse 
B-field in the central channel of the jet and longitudinal B-field at the edges is seen on epoch 2008.70.
{\it Right panel:} Brightness temperature changes along the jet in 2155$-$152 at 15.4~GHz. Filled blue circles
are the measured values. Open red circles are the predicted brightness temperatures from shock-in-jet model 
with adiabatic losses dominating the radio emission assuming predominantly transverse magnetic field 
orientation to the local jet direction. The errors are given at $3\sigma$ level.
}
\label{fig4}
\end{figure}

Using the parsec-scale spectral index information (Fig.~\ref{fig3}, right) taken from the I-maps
at 8.1 and 15.4~GHz  (Fig.~\ref{fig3}, left) together with the results from model fitting, and also 
assuming magnetic field to be perpendicular to the jet, we have calculated the values of 
$T^{\rm mod}_{\rm\,\,\,b}$ and $T^{\rm obs}_{\rm\,\,\,b}$. The brightness temperatures predicted by the model 
agree well with those from the data for all the components (Fig.~\ref{fig4}, right), except the optically 
thin J2 feature ($\alpha_{J2}=-0.75$) located at 3.36~mas from the core, for which 
$\zeta=T^{\rm obs}_{\rm\,\,\,b}/T^{\rm mod}_{\rm\,\,\,b}\approx20$. There are two possible ways to explain 
this discrepancy: 

\noindent
(i) The J2 component is the region of the outflow where the particles are reaccelerated through the 
interaction with the surrounding medium; an evidence for this is the formation of a shear layer as 
is described in the decelerating jet model \cite{Laing96}: in the central channel, where magnetic 
field is predominantly tranverse, plasma moves with relativistic speed but slows down at the edges due to 
interaction with an ambient medium making a sheath of longitudinal magnetic field registered on
polarization images (Fig.~\ref{fig4}, left).

\noindent
(ii) The Doppler factor varies in the J2 region due to changes of the speed or orientation of the jet.
The apparent speed $\beta_{\rm app\,\,J5}=(10.8\pm1.3)c$ is determined from the MOJAVE kinematics 
\cite{Kellermann04,Lister08}. The J2 component is quasi-stationary (during a period of 5.2\,yr
centroid position was wiggling within 0.3\,mas in non-radial direction). 
Then assuming the viewing angle $\theta_{\rm max}\approx5.3^\circ$ ($\theta\sim1/\beta$)
to be constant and applying analytical adiabatic approximation \cite{Baum97} for a relativistic jet with
predominantly transverse B-field,
we determined the change of speed $\Delta\beta\le0.002$ between J5 and J2 regions. Therefore, the detected 
excess of the Doppler factor $\delta_{\rm J2}/\delta_{J5}=\zeta^{1/(3-\alpha_{\rm J2})}=2.22$ can be explained by 
changing the viewing angle that becomes less than the opening angle of the jet $\phi\approx1^\circ$.


\clearpage
\newpage

\section{Conclusion}

The measured sizes and brightness temperatures of the VLBI components in parsec-scale
jets in radio galaxy 1128$-$047 and quasar 2155$-$047 are found to be consistent with emission
from thin relativistic shocks dominated by adiabatic energy losses. The bright distinct features
in these sources may indeed be a collection of relativistic shocks developing and expanding
down the flow. Applying the shock-in-jet model to the VLBI observations we predict the predominantly
transverse direction of the intrinsic magnetic field associated with the central channel of the 
jet in 1128$-$047. Discrepancy between the model and measured brightness temperatures for the bright jet 
component, located at $\sim$3.4~mas from the core in 2155$-$152, can be explained either by 
reacceleration of the particles due to the interaction with the ambient medium or by changing the 
viewing angle that becomes nearly zero.

\end{document}